\documentstyle[11pt,epsfig]{article}

\begin{document}

\begin{center}

\Large LHC(CMS) SUSY discovery potential for the case \\
of nonuniversal gaugino masses$^1$\\

\bigskip 

\normalsize

N.V.~Krasnikov\\
Institute for Nuclear Research RAS, Moscow 117312, Russia\\ 
E-mail: Nikolai.Krasnikov@cern.ch

\bigskip

S.I.~Bityukov\\
Institute for High Energy Physics, Protvino, Russia\\
E-mail: Serguei.Bitioukov@cern.ch \\ 

\end{center}


We investigate squark and gluino  pair production
at LHC(CMS) for the case 
of nonuniversal gaugino masses. Visibility of signal by an excess
over SM background in $(n \geq 2)jets + E^{miss}_T + (m \geq leptons)$ events
depends rather strongly on the relation between LSP, gluino and 
squark masses and it decreases with the increase of LSP mass. 
For relatively heavy LSP mass close to squark or gluino masses 
it is possible to detect SUSY  
for $(m_{\tilde{q}}, m_{\tilde{g}}) \leq (1 - 1.5)$ TeV.

\vspace{0.5cm}

\noindent
\rule{3cm}{0.5pt}\\
$^1$~~To appear in the Proceedings of SUSY01, Dubna (Russia), June 2001 \\

\bigskip

\section{Introduction}
One of the LHC supergoals is the discovery of the supersymmetry.  
In ref.\cite{1} (see, also references \cite{2}) the LHC(CMS) SUSY discovery 
potential has been investigated within the minimal SUGRA-MSSM model  \cite{3} 
where all sparticle masses are determined mainly by two parameters: 
$m_0$ (common squark and slepton mass at GUT scale) and 
$m_{1 \over 2}$ (common gaugino mass at GUT scale). 
The signature used for the search for squarks and gluino  at LHC is 
$(m \geq 0)$ leptons + $(n \geq 2)jets + E^{miss}_T$ events. 
The conclusion of ref.~\cite{1}  is that LHC(CMS) is able 
to detect squarks and gluino  with masses up to 
(2 - 2.5)~TeV.  

Despite the simplicity of the  SUGRA-MSSM model  it is a very particular
model and  we can expect that real 
sparticle masses can differ in a drastic way 
from sparticle masses pattern of SUGRA-MSSM model due to many reasons,
see for instance refs.~\cite{4,5,6}.
Therefore, it is more appropriate to investigate the LHC SUSY discovery 
potential in a model-independent way.

\section{Results}
This talk is based on our papers \cite{7} where the reader 
can find additional information.
Our main conclusion is  that LHC SUSY  discovery potential depends rather 
strongly on the relation between  
squark, gluino and LSP masses and it decreases with the 
increase of the LSP mass. For LSP mass close to squark or gluino masses it 
is possible to detect SUSY at LHC for squark and gluino masses up to 
$(1 - 1.5)~TeV$.
Our results on LHC(CMS) SUSY discovery potential are presented in 
Figs.1-2.

\begin{figure}[t]
\epsfxsize=20pc 
\epsfysize=19pc 
\epsfbox{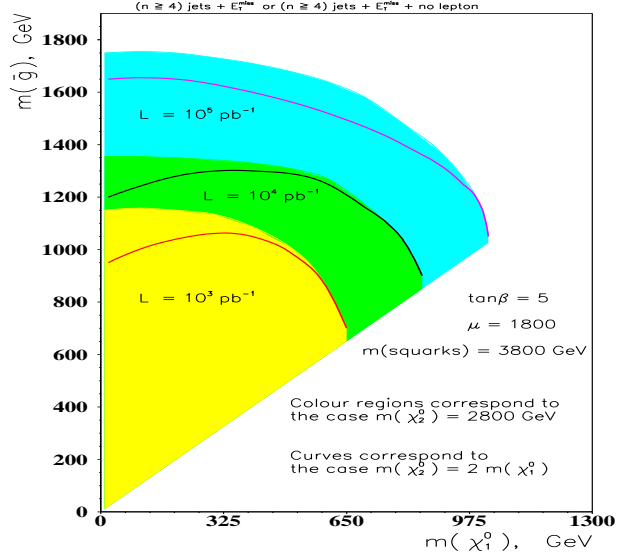} 
\caption{ CMS discovery potential for different values
of $m_{\tilde \chi_1^0}$ and $m_{\tilde g}$ in the case of 
$m_{\tilde g}~>~m_{\tilde \chi^0_1}$~~($m_{\tilde q} \gg m_{\tilde g}$). 
\label{fig:fig1}}
\end{figure}

\begin{figure}[t]
\epsfxsize=20pc 
\epsfysize=19pc 
\epsfbox{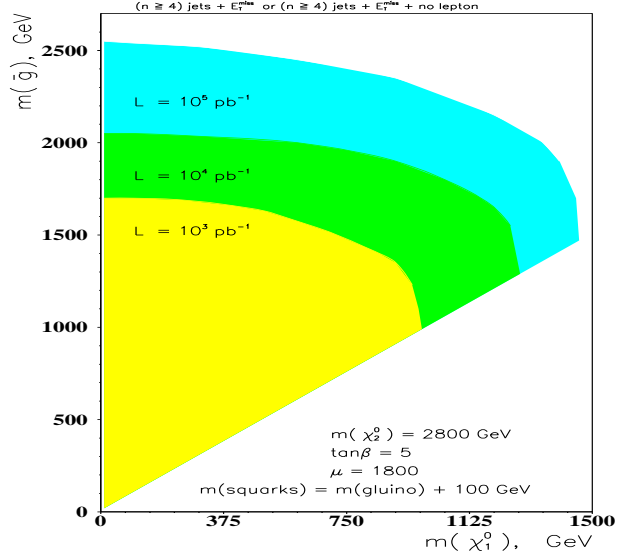} 
\caption{CMS discovery potential for different values
of $m_{\tilde \chi_1^0}$ and $m_{\tilde g}$ in the case of 
$m_{\tilde g}~>~m_{\tilde \chi^0_1}$~~
($m_{\tilde q} = m_{\tilde g}$ + 100~GeV). 
\label{fig:fig2}}
\end{figure}

\section*{Acknowledgments}
We are  indebted to 
the participants of Daniel Denegri working group on physics 
simulations at LHC for useful comments. 
This work has been supported by INTAS-CERN 377 and RFFI grant
99-01-00091.

\end{document}